\documentclass[nofootinbib,12pt]{revtex4-1}
\pdfoutput=1

\makeatletter
\renewcommand{\p@subsection}{}
\renewcommand{\p@subsubsection}{}
\makeatother

\usepackage{graphicx}
\usepackage{amsmath}
\usepackage{amssymb}
\usepackage[mathscr]{euscript}
\usepackage{enumerate}
\usepackage{hyperref}

\hypersetup{
	colorlinks=true,
	linktoc=page,
	linkcolor=blue,
	citecolor=blue,
	urlcolor=blue
	}

\DeclareMathOperator{\Tr}{Tr}
\newcommand{\Sc}[0]{ {\mathscr S} }
\newcommand{\ScOP}[0]{ {\mathscr S}^{(1)} }
\newcommand{\Ssc}[0]{ {\mathscr S}^{(\wedge)} }
\newcommand{\Area}[0]{ \text{Area} }
\newcommand{\bdiamond}[0]{ \blacklozenge }

\begin{document}

\title{Deriving the First Law of Black Hole Thermodynamics without Entanglement}

\author{William R. Kelly}
\email{wkelly@physics.ucsb.edu}
\affiliation{University of California at Santa Barbara, Santa Barbara, CA 93106, USA}
\date{\today}

\begin{abstract}
In AdS/CFT, how is the bulk first law realized in the boundary CFT?  Recently, Faulkner et al. showed that in certain holographic contexts, the bulk first law has a precise microscopic interpretation as a first law of entanglement entropy in the boundary theory.  However, the bulk can also satisfy a first law when the boundary density matrix is pure, i.e. in the absence of entanglement with other degrees of freedom.  In this note we argue that the bulk first law should generally be understood in terms of a particular coarse-graining of the boundary theory.  We use geons, or single-exterior black holes, as a testing ground for this idea.  Our main result is that for a class of small perturbations to these spacetimes the Wald entropy agrees to first order with the one-point entropy, a coarse-grained entropy recently proposed by Kelly and Wall.  This result also extends the regime over which the one-point entropy is known to be equal to the causal holographic information of Hubeny and Rangamani.
\end{abstract}

\maketitle



\section{Introduction} \label{sec:intro}

The Wald-Iyer theorem~\cite{Wald:1993nt,*Iyer:1994ys} establishes that the first law of black hole thermodynamics~\cite{Bardeen:1973gs} is a general consequence of diffeomorphism invariance.  In the context of AdS/CFT, it has been shown by Faulkner et al.~\cite{Faulkner:2013ica} that a special case of the Wald-Iyer theorem has a precise microscopic interpretation as the `first law of entanglement entropy'~\cite{Blanco:2013joa}.  This insight turned out to be very powerful, as it led to a derivation of the linearized Einstein equation~\cite{Faulkner:2013ica} from the Ryu-Takayanagi formula~\cite{Ryu:2006bv,*Ryu:2006ef} (see also~\cite{Hammersley:2006cp,*Hammersley:2007ab,Bilson:2008ab,Nozaki:2013vta,Lashkari:2013koa,Bhattacharya:2013bna}).\footnote{Note that the linearized EOM can also be derived (under a different set of assumptions) from conformal invariance (see~\cite{Kabat:2012hp}).}  Subsequent work extended this derivation to include universal coupling to matter~\cite{Swingle:2014uza} (with an additional assumption argued for in~\cite{Faulkner:2013ana}).

Given this recent success, it seems both interesting and important to answer the question `What is the holographic dual of the Wald-Iyer theorem?'.  In light of the previous paragraph one might naively guess that the Wald-Iyer theorem is the bulk dual of the first law of entanglement entropy, however, as we will show below, this guess is incorrect.  Instead we will argue that the Wald-Iyer theorem is dual to a coarse-grained first law.  More precisely, we will prove that for a certain class of states defined in section~\ref{sec:CG}
\begin{align}\label{eq:ScOPeqSW}
\delta S_W = \delta \ScOP.
\end{align}
Here $S_W$ is the Wald entropy, $\ScOP$ is the one-point entropy of~\cite{Kelly:2013aja}, and $\delta$ is a variation which acts infinitesimally on both the bulk spacetime and the boundary density matrix.  The one-point entropy (which we define in section~\ref{sec:review}) is a coarse-grained measure of information that is only sensitive to the expectation value of local operators (i.e. one-point functions) within a boundary causal domain of dependence.  Our main result is that~\eqref{eq:ScOPeqSW} holds even for pure states, for which the Wald entropy is not a measure of entanglement of the associated CFT state.

For many states, including the AdS-Rindler state considered in~\cite{Faulkner:2013ica},~\eqref{eq:ScOPeqSW} does reduce to the first law of entanglement entropy $\delta S_W = \delta S$, where $S$ is the von Neumann entropy.  Still, there are two reasons why our interpretation of the Wald entropy as a coarse-grained entropy is useful.

First, there are other states for which $\delta S_W \ne \delta S$ but~\eqref{eq:ScOPeqSW} continues to hold.  Examples of such states are 
\begin{itemize}

\item
topological-geon/single-exterior black holes~\cite{Sorkin:1986}

\item
the ``B-states" of~\cite{Hartman:2013qma}, which model a CFT excited state after a global quench (see~\cite{Calabrese:2005in})

\item
black hole microstates of either the fuzzy (see e.g.~\cite{Mathur:2005zp}) or fiery~\cite{Braunstein:2009my,Almheiri:2012rt} persuasion

\item
the late time limit of a collapsed black hole.

\end{itemize}
What these states have in common is that, even though they are dual to pure (or nearly pure) CFT states, they each have a bulk region which resembles a black hole, including obeying a thermodynamic first law.\footnote{See also~\cite{Balasubramanian:2014hda}.}   This latter behavior is captured by~\eqref{eq:ScOPeqSW}.

Second, a corollary of our result and~\cite{Faulkner:2013ica} is that the linearized gravitational equations of motion can also be derived from~\eqref{eq:ScOPeqSW}.  This observation suggests that it might be possible to derive gravitational equations of motion from a coarse graining of the microscopic degrees of freedom, in the spirit of~\cite{Jacobson:1995ab}.  This proposal could be tested by deriving the linearized equations using states for which $\ScOP \ne S $ or by checking to see if~\eqref{eq:ScOPeqSW} continues to hold beyond linear order.

\begin{figure}
\includegraphics[width=0.4 \textwidth]{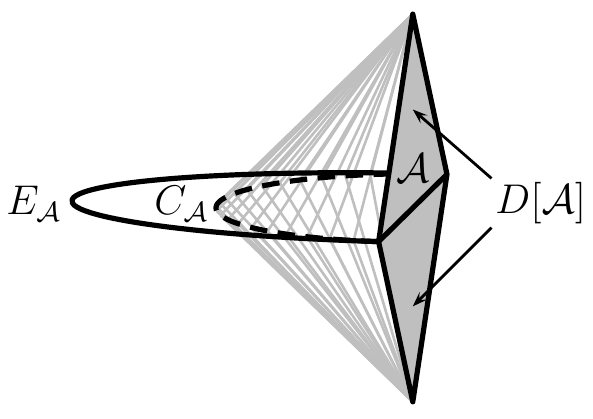} 
\caption{A sketch of a boundary region $\cal A$, its associated domain of dependence $D[{\cal A}]$, the causal information surface $C_{\cal A}$ and the RT/HRT surface $E_{\cal A}$~\cite{Ryu:2006bv,*Ryu:2006ef,Hubeny:2007xt}.  $D[{\cal A}]$ lies on the AdS boundary while $C_{\cal A}$ and $E_{\cal A}$ extend into the bulk spacetime.  The wedge shaped region enclosed by $D[{\cal A}]$ along with the bulk past and future horizons of $D[{\cal A}]$ (gray lines) is called the causal wedge of $\cal A$ and denoted $\bdiamond_{\cal A}$.}
\label{fig:causal}
\end{figure}

Equation~\eqref{eq:ScOPeqSW} also has implications for the proposal of~\cite{Kelly:2013aja}.  In~\cite{Kelly:2013aja} it was conjectured that, in the Einstein gravity limit, the one-point entropy could be computed from the `Ryu-Takayanagi'-like formula
\begin{align} \label{eq:SeqCHI}
\ScOP(\rho_{\cal A}) = \frac{\text{Area}[C_{\cal A}]}{4G} =: \chi_{\cal A}.
\end{align}
Here $\rho_{\cal A}$ is the reduced density matrix associated with a CFT region $\cal A$, $C_{\cal A}$ is the intersection of the past and future horizons of $D[{\cal A}]$, $D[{\cal A}]$ is the boundary domain of dependence of $\cal A$, and $\chi$ is the causal holographic information (CHI) of~\cite{Hubeny:2012wa} defined above in~\eqref{eq:SeqCHI}
.\footnote{The proposal as stated applies only to Einstein-Hilbert gravity, but there is a natural generalization to higher derivative theories of gravity by replacing the $\text{Area}$ functional with the entropy functional of~\cite{Bhattacharyya:2013gra,Dong:2013qoa,Camps:2013zua,Bhattacharyya:2014yga}.  In this note we will only be interested in cases for which this entropy functional reduces to the Wald entropy.}  Since~\eqref{eq:ScOPeqSW} is a first variation of~\eqref{eq:SeqCHI} for a class of special states, our proof of~\eqref{eq:ScOPeqSW} provides new evidence for the conjecture $\ScOP = \chi$. 

The organization of the rest of the paper is as follows.  In section~\ref{sec:review} we review the definition of $\ScOP$ and briefly state some of the motivation for~\eqref{eq:SeqCHI}.  In section~\ref{sec:CG} we prove our main result $\delta \ScOP = \delta S_W$ and provide examples of states that satisfy the assumptions of our proof.  In section~\ref{sec:Disc} we summarize our results and comment on their relationship to the related work of~\cite{Balasubramanian:2013lsa,Myers:2014jia,Czech:2014wka,Balasubramanian:2014sra}.  In Appendix~\ref{app:tests} we outline a strategy for testing~\eqref{eq:SeqCHI} non-perturbatively.

\section{The one point entropy $\ScOP$} \label{sec:review}

In this section we briefly define and motivate the one-point entropy $\ScOP$, we refer the reader to~\cite{Kelly:2013aja} for additional details.  The one-point entropy is defined as
\begin{align} \label{eq:OnePointDef}
\ScOP(\rho_{\cal A}) =  \underset{\tau_{\cal A} \in T_{\cal A}}{\text{lub}} S(\tau_{\cal A}),
\end{align}
where $\rho_{\cal A}$ is the reduced density matrix associated with a spacelike region $\cal A$ of the CFT, $S(\tau_{\cal A}) := -\Tr[\tau_{\cal A}\log(\tau_{\cal A})]$ is the von Neumann entropy, and `$\text{lub}$' stands for the least upper bound, in this case over the states $ T_{\cal A}$.  Here, $T_{\cal A}$ is the set of all states $\tau_{\cal A}$ which satisfy
\begin{align} \label{eq:constraints}
\Tr[{\cal O}(x) \tau_{\cal A}] = \Tr[{\cal O}(x) \rho_{\cal A}], \qquad x \in D[{\cal A}],
\end{align}
for all local, gauge invariant CFT operators ${\cal O}(x)$.  In words, $\ScOP(\rho_{\cal A})$ is the least upper bound of the von Neumann entropy of all state $\tau_{\cal A}$ which reproduce the one-point functions of all local operators in the domain of dependence $D[{\cal A}]$.\footnote{See~\cite{GellMann:2006uj} for a non-holographic application of this type of coarse-graining.}

Heuristically, we might imagine an experimental physicist performing all local measurements in $D[{\cal A}]$ and trying to estimate the state $\rho_{\cal A}$ based only on this data.  Having no other information at her disposal, this experimentalist would be justified in assigning equal probabilities to any state that reproduces her measurements.  The entropy of the resulting ensemble is precisely $\ScOP(\rho_{\cal A})$.

One feature of~\eqref{eq:SeqCHI} is that it implies that $\Area[C_{\cal A}]$ can be expressed as a function of local measurements in $D[{\cal A}]$.  In the large $N$ limit CFT correlation functions factorize and local measurements are roughly equivalent to measuring all correlators at leading order in a $1/N$ expansion.  This intuition along with the bulk reconstruction literature~\cite{Balasubramanian:1998sn,Balasubramanian:1998de,Banks:1998dd,Bena:1999jv,Hamilton:2005ju,Hamilton:2006az,Kabat:2011rz,Heemskerk:2012mn,Morrison:2014jha} suggests that, at least perturbatively, the one-point functions are sufficient to construct the classical spacetime up to $C_{\cal A}$. 

An important implication of~\eqref{eq:OnePointDef} is that whenever the modular Hamiltonian of $\rho$ is local, we must have $\ScOP(\rho) = S(\rho)$.  Recall that the modular Hamiltonian $H$ is defined for any positive definite $\rho$ by the relation
\begin{align} \label{eq:modHdef}
\rho = Z^{-1} \exp(-H),
\end{align}
where $Z = \Tr[\exp(-H)]$ and $H$ is generically a complicated non-local operator.  If $H$ is local (or more precisely the integral of a local operator) then $\left< H\right>_{\tau_{\cal A}}$ is fixed by the constraints~\eqref{eq:constraints}.  It is a standard result of thermodynamics that $\rho$ maximizes the von Neumann entropy subject to the constraint of fixed $\left< H \right>$, therefore~\eqref{eq:OnePointDef} reduces to $\ScOP(\rho) = S(\rho)$.  In AdS/CFT, $H$ is local only for very special states, such as stationary black holes and AdS-Rindler, and in all such cases we find that the minimum area surface $E_{\cal A}$ picked out by the Ryu-Takayanagi (RT) conjecture~\cite{Ryu:2006bv,*Ryu:2006ef} (or equivalently the minimum area, extremal surface picked out by the Hubeny-Rangamani-Takayanagi (HRT) conjecture~\cite{Hubeny:2007xt}) and the causal information surface $C_{\cal A}$ coincide~\cite{Hubeny:2012wa}.  The RT/HRT conjectures state that $S= \Area(E_{\cal A})/4G$, which implies that for these special states $\chi =S$.  Slightly abusing the standard terminology, we will refer to states of this kind as `thermal' even when $H$ is not the generator of time translations.

For any density matrix of the form~\eqref{eq:modHdef} a simple calculation yields the first law of entanglement entropy
\begin{align} \label{eq:deltaS}
S(\rho+\delta\rho) = S(\rho) + \Tr[\delta\rho H] + O(\delta\rho^2).
\end{align}
We will use this identity frequently below.

Finally, if we assume that CFT states with semi-classical bulk geometries are appropriately generic (see~\eqref{eq:assumption}), then~\eqref{eq:SeqCHI} reduces to a statement about the classical equations of motion which in principle is testable.  The interested reader may consult Appendix~\ref{app:tests} for the details.

\section{A Proof of $\delta S_W = \delta \ScOP$}\label{sec:CG}

In this section we prove~\eqref{eq:ScOPeqSW} under a set of assumptions.  We then provide examples of states satisfying those assumptions.

\subsection{The General Case} \label{sec:proof}

The assumptions for our proof of~\eqref{eq:ScOPeqSW} are as follows.  Let $\cal A$ be a spacelike region of the CFT (possibly an entire Cauchy surface) and let $\rho_{\cal A}$ be the reduced density matrix on $\cal A$.  We assume that:
\begin{enumerate}[(I)]

\item \label{assume:classical}
The dual bulk state is well approximated by a semiclassical bulk geometry, at least up to an order Planck length distance from the boundary of the causal wedge $\bdiamond_{\cal A}$ (see Fig.~\ref{fig:causal}).  The rest of the bulk need not be semiclassical.

\item \label{assume:stationary}
The interior of $\bdiamond_{\cal A}$ is stationary with Killing vector $t$ and is isometric to the interior of another spacetime region $\hat \bdiamond_{\cal A}$ which has a bifurcate Killing horizon as its boundary.  Let $\xi$ be a Killing vector in the interior of $\bdiamond_{\cal A}, \hat \bdiamond_{\cal A}$, which vanishes on the bifurcation surface of $\hat \bdiamond_{\cal A}$.  We fix the normalization of $t$ and $\xi$ by requiring that, at the conformal boundary, $t\cdot t = -1$ and $\xi \cdot t = -1$.

\item \label{assume:onepoint}
The one-point functions of $\rho_{\cal A}$ are identical to the one-point functions of a state $\rho_{th} $, where $\rho_{th}$ is of the form
\begin{align} \label{eq:rhoth}
\rho_{th} = Z^{-1} \exp(-H_{th}), \qquad H_{th} =  \int_\Sigma n^a T_{ab} \xi^b.
\end{align}
Here $Z = \Tr[\exp(-H_{th})]$, $\Sigma$ is a Cauchy surface of the boundary region $D[{\cal A}]$, $n^a$ is the associated unit normal, $T_{ab}$ is the boundary stress tensor, and $\xi^a$ is the pullback of $\xi$ to the conformal boundary.

\end{enumerate}

Assumptions~\eqref{assume:classical} and~\eqref{assume:stationary} are needed so that we may invoke the Wald-Iyer theorem.  We were careful to word~\eqref{assume:stationary} so as not to require that the boundary of $\bdiamond_{\cal A}$ be a Killing horizon.  This distinction will be important later when we consider geometries like the $\mathbb{RP}_n$ geon which have an exterior region that is isometric to a stationary black hole, but do not have a bifurcate Killing horizon.\footnote{Note that the surface integral often used to calculate the Wald entropy arises from integrating a total divergence over a bulk Cauchy surface.  For this reason the Wald entropy, properly defined, is the same on $\bdiamond_{\cal A}$ and $\hat \bdiamond_{\cal A}$, which is why assumption~\eqref{assume:stationary} is sufficient for our purposes. \label{foot:BT} }

Assumption~\eqref{assume:onepoint} expresses the intuition that stationary geometries are consistent with thermal states.  Known examples suggest that~\eqref{assume:onepoint} holds if and only if~\eqref{assume:classical} and~\eqref{assume:stationary} also hold, which implies that it may be possible to derive~\eqref{assume:onepoint} from~\eqref{assume:classical} and~\eqref{assume:stationary}.  It would be an improvement to eliminate~\eqref{assume:onepoint}, but for now we will take it as an assumption and argue that it is satisfied for the states listed in the introduction.

For simplicity we have not considered charged black holes, but it would be straightforward to do so using the results of~\cite{Gao:2001ut}.  We now begin the proof.

{\bf Theorem: Assumptions~\eqref{assume:classical}-\eqref{assume:onepoint} imply $ \delta S_W = \delta \ScOP$, where $\delta$ is a variation that acts infinitesimally both on the boundary state $\rho_{\cal A}$ and the bulk geometry.}

Our strategy will be to calculate $\delta S_W$ and $\delta \ScOP$ separately and compare the answers.  We begin with $\delta S_W$.  By assumptions~\eqref{assume:classical} and~\eqref{assume:stationary}  we may invoke the Wald-Iyer theorem which states that
\begin{align}\label{eq:WaldTheorem}
\delta S_W = \delta {\cal H},
\end{align}
where ${\cal H}$ is the canonical charge associated with the Killing vector $\xi$.\footnote{${\cal H}$ is defined by the differential equation $\delta{\cal H} = \omega(\delta \phi, \pounds_\xi \phi)$, where $\omega$ is the symplectic structure, $\pounds_\xi$ is the Lie derivative along $\xi$, and $\phi$ represents the metric and any other field content of the theory.  We have chosen conventions which set the temperature to unity.}   It has been shown explicitly~\cite{Papadimitriou:2005ii,Hollands:2005wt} (or more generally in~\cite{Hollands:2005ya}) that $\cal H$ is equal to the holographic charge associated with $\xi$ up to a term that is constant on the space of solutions, i.e.
\begin{align} \label{eq:Hth}
\left<H_{th}\right> := \int_\Sigma n^a T_{ab} \xi^b = {\cal H} + c,
\end{align}
where $T_{ab}$ is the holographic stress tensor computed using the counter term subtraction prescription of~\cite{Henningson:1998gx,Balasubramanian:1999re}.  Since $c$ is a constant on the space of solutions, it will vanish when we take the variational derivative with respect to the bulk solution, so we may rewrite~\eqref{eq:WaldTheorem} as
\begin{align} \label{eq:delSW}
\delta S_W =  \delta\left<H_{th}\right>.
\end{align}

Now we turn to calculating $\delta\ScOP$.  Let the variation of the bulk geometry considered above correspond to a variation of the density matrix
\begin{align}
\rho_{\cal A}\to \rho_{\cal A}+\delta\rho.
\end{align}
We now wish to compute 
\begin{align}
\delta\ScOP = \delta\ScOP(\rho_{\cal A}+\delta\rho) - \ScOP(\rho_{\cal A})  + O(\delta\rho^2).
\end{align}
It turns out to be useful to consider the family of states $\rho_{\cal A}+\alpha \, \delta\rho$, where $\alpha$ is an arbitrary constant.  Recall from section~\ref{sec:review} that $\ScOP$ is calculated by maximizing the entropy over states which satisfy a constraint of the form~\eqref{eq:constraints}.  By assumption~\eqref{assume:onepoint}, $\rho_{\cal A} +\alpha \, \delta\rho$ must have identical one-point functions to $ \rho_{th} + \alpha \, \delta\rho $, therefore
\begin{align} \label{eq:Sbound}
\ScOP(\rho_{\cal A} + \alpha \, \delta \rho) = \ScOP(\rho_{th} +\alpha \, \delta\rho) \ge S(\rho_{th} +\alpha \, \delta\rho),
\end{align}
where the last inequality follows from the definition~\eqref{eq:OnePointDef}.

Also by assumption~\eqref{assume:onepoint} we have $\ScOP(\rho_{th}) = S(\rho_{th})$ because $\rho_{th}$ has a local modular Hamiltonian (by the argument given just below~\eqref{eq:modHdef}).  Inserting this relation into~\eqref{eq:Sbound} and using~\eqref{eq:deltaS} gives
\begin{align}
\alpha \left( \delta\ScOP- \Tr[ \delta\rho \,  H_{th}] \right) + O(\alpha^2) \ge 0,
\end{align}
This inequality must hold for arbitrary $\alpha$, therefore the term in parenthesis vanishes,\footnote{Thanks to Aron Wall for pointing out that my original argument could be considerably simplified.} and
\begin{align} \label{eq:delSOP}
\delta\ScOP = \delta\left<H_{th}\right>.
\end{align}
Comparing~\eqref{eq:delSW} and~\eqref{eq:delSOP} we see that the proof is complete.  We now prove a corollary which will be used below.

{\bf Corollary: Under the same assumptions as above, $\delta S_W =  \delta S(\rho_{\cal A}) $ if and only if $\rho_{\cal A} = \rho_{th} $ for $\rho_{th}$ as defined in~\eqref{eq:rhoth}.}  

If $\rho_{\cal A} = \rho_{th} $ then it follows immediately from~\eqref{eq:deltaS} and~\eqref{eq:delSW} that
\begin{align}
\delta S = \delta \left< H_{th}\right> = \delta S_W.
\end{align}
Conversely, say that $\delta S_W =  \delta S$ for all $\delta\rho$.  It then also follows from~\eqref{eq:deltaS} and~\eqref{eq:delSW} that
\begin{align} \label{eq:HA}
\Tr[\delta\rho \, H_{th}] = \Tr[\delta\rho \, H_{\cal A}] ,
\end{align}
where $H_{\cal A}$ is the modular Hamiltonian of $\rho_{\cal A}$.  But~\eqref{eq:HA} can only hold for arbitrary $\delta\rho$ if $H_{\cal A} = H_{th}$, which implies that $\rho_{\cal A} = \rho_{th}$.  This completes our proof of the corollary.

\subsection{Stationary Examples}

\begin{figure}
\includegraphics[width=0.6 \textwidth]{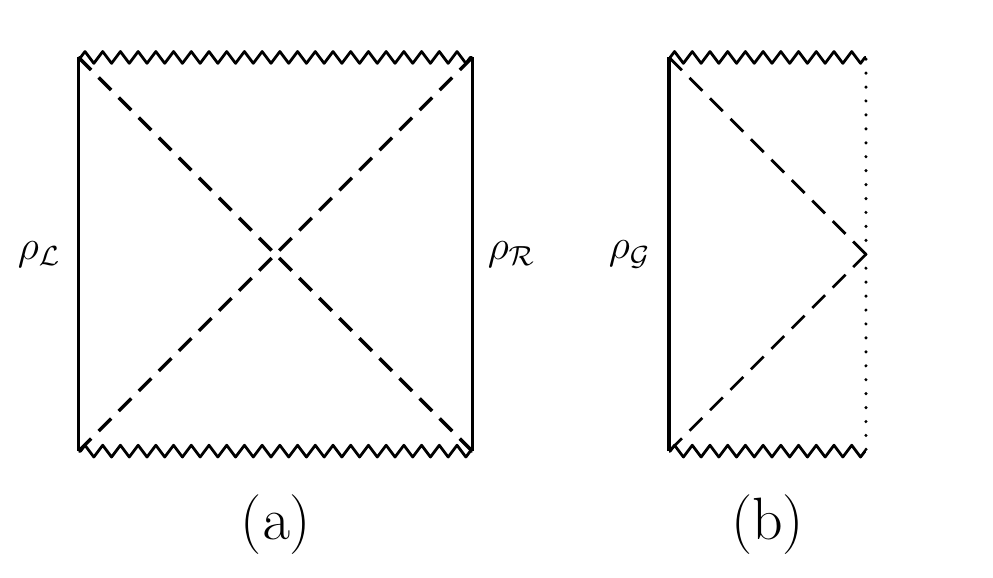} 
\caption{(a) A causal diagram AdS Schwarzschild.  The reduced density matrix $\rho_{\cal L}$ is an example of a state for which $\delta S_W = \delta S$.  (b) A causal diagram of the geon spacetime described in the text.}
\label{fig:BH}
\end{figure}

There are many examples of states which satisfy assumptions~\eqref{assume:classical}-\eqref{assume:onepoint}.  One natural example comes from the thermofield double state, which is dual to the two sided AdS-Schwarzschild geometry~\cite{Maldacena:2001kr}.  If we let $\cal L$ be a Cauchy surface of the left boundary (see Fig.~\ref{fig:BH}(a)), then $\bdiamond_{\cal L}$ is the exterior region of AdS-Schwarzschild, which satisfies~\eqref{assume:classical} and~\eqref{assume:stationary}.  The reduced density matrix of the left asymptotic region, $\rho_{\cal L}$, is already of the form~\eqref{eq:rhoth}, therefore~\eqref{assume:onepoint} is satisfied.  Additionally, $\rho_{\cal L}$ satisfies the condition of the corollary, therefore $\delta\ScOP = \delta S$ and~\eqref{eq:ScOPeqSW} reduces to the first law of entanglement entropy.

As promised in the introduction we will now show that there exist states for which $\delta\ScOP \ne \delta S$ but~\eqref{eq:ScOPeqSW} still holds.  By the corollary proved in section~\ref{sec:proof} this amounts to showing that there exists a state satisfying assumptions~\eqref{assume:classical}-\eqref{assume:onepoint} for a density matrix $\rho_{\cal A}$ that is \emph{not} a thermal state of the form~\eqref{eq:rhoth}. 

In fact there are large classes of such states.  One class of examples are known as topological geons~\cite{Sorkin:1986}.  A simple example of a geon is the (AdS) $\mathbb{RP}_n$ geon (see e.g.~\cite{Friedman:1993ty}).  This solution can be constructed from a $t=0$ Cauchy slice of maximally extended AdS-Schwarzschild by taking a $\mathbb{Z}_2$ quotient about the bifurcation surface $\cal B$ and identifying antipodal points on $\cal B$.  The resulting surface has a topology $\mathbb{RP}_n$ where $n$ is the dimension of the Cauchy surface, hence the name.  The maximal evolution of this new surface is a smooth spacetime with one asymptotic region  (see Fig.~\ref{fig:BH}(b)).

Let $\cal G$ be a Cauchy surface of the geon boundary with associated density matrix $\rho_{\cal G}$.  By construction the interior of $\bdiamond_{\cal G}$ is identical to the exterior of the AdS-Schwarzschild black hole, therefore the CFT state $\rho_{\cal G}$ satisfies assumptions~\eqref{assume:classical} and~\eqref{assume:stationary}.  Furthermore, by the usual AdS/CFT dictionary the one-point functions of $\rho_{\cal G}$ are identical to the one point functions of $\rho_{\cal L}$, the density matrix of the left boundary of AdS-Schwarzschild.\footnote{Modulo an issue related to choice of conformal frame, which is non-trivial in the presence of a conformal anomaly (see~\cite{Skenderis:2009ju}).  However, this anomaly term only modifies $H_{th}$ by a constant $c$ as in~\eqref{eq:Hth}, which we have already accounted for.  Thanks to Kostas Skenderis for pointing this out to me.}  So the state $\rho_{\cal G}$ also satisfies assumption~\eqref{assume:onepoint}.

It only remains to show that $\rho_{\cal G} \ne \rho_{\cal L}$.  This is most easily seen by calculating the entropy of both states.  The entropy of $\rho_{\cal L}$ is given by $S(\rho_{\cal L}) = S_W \sim N^2$.  The geon geometry, on the other hand, has vanishing Ryu-Takayanagi entropy, which implies that the entropy $\rho_{\cal G}$ is parametrically smaller than $N^2$.   Other arguments, given in~\cite{Maldacena:2001kr} and explained in detail in~\cite{Skenderis:2009ju} (see also~\cite{Louko:1998hc,*Louko:1999xb,*Louko:2000tp}) indicate that $\rho_{\cal G}$ can be chosen to be a pure state.\footnote{Up to this point we had not completely specified $\rho_{\cal G}$.}  Therefore, by the corollary proved in section~\ref{sec:CG}, $\rho_{\cal G}$ is a state for which
\begin{align} \label{eq:nontrivial}
\delta S_W = \delta \ScOP \ne \delta S.
\end{align}

As mentioned in the introduction, another state satisfying assumptions~\eqref{assume:classical}-\eqref{assume:onepoint} is the B-state constructed in~\cite{Calabrese:2005in} and studied holographically in~\cite{Hartman:2013qma}.  This state is a pure CFT state meant to model a global quench, in which the Hamiltonian of the theory is changed abruptly.  Hartman and Maldacena~\cite{Hartman:2013qma} argued that bulk geometry of the B-state can be obtained by slicing the maximally extended AdS-Schwarzschild geometry in half and terminating the spacetime in an end of the world brane.  They then used the Ryu-Takayanagi proposal to reproduce the time evolution of the entanglement entropy calculated in the field theory by Calabrese and Cardy~\cite{Calabrese:2005in}.

It follows immediately from the construction described above that the B-state spacetime has a conformal diagram like Fig.~\ref{fig:BH}(b) and satisfies~\eqref{assume:classical}-\eqref{assume:onepoint} by the same arguments as in the geon case.  Since the B-state is pure,~\eqref{eq:nontrivial} also follows just as for the geon states.

As our last example we consider the firewall~\cite{Braunstein:2009my,Almheiri:2012rt} and fuzzball (see~\cite{Mathur:2005zp}) proposals.  Both proposals predict that black hole states are ensembles of pure states each of which matches the classical geometry from asymptotic infinity up to a few Planck lengths from the horizon, and beyond this stretched horizon the semiclassical description fails.  These microstates---which have been explicitly constructed for certain external black holes (see~\cite{Mathur:2008nj,Balasubramanian:2008da,Chowdhury:2010ct} for a review)---provide another example of pure states which satisfy~\eqref{assume:classical}-\eqref{assume:onepoint}.

\subsection{Collapsed black holes} \label{sec:Collapse}

Another interesting class of pure (or nearly pure) state black holes are given by black holes formed from collapse.  States of this kind satisfy~\eqref{assume:classical} but not~\eqref{assume:stationary} because the resulting geometry is not stationary.  As a result, we cannot directly apply the theorem of section~\ref{sec:CG} to these states.  However, we can make some progress if we consider collapsed black holes that asymptote to stationary black holes at late times.

Let $\rho_{\cal C}$ be a state describing a black hole formed from collapse that settles down to a stationary black hole defined on a Cauchy surface $\cal C$.  Let $\rho_{th}$ be a thermal state of the form~\eqref{eq:rhoth} dual to that stationary black hole, and assume that the one-point functions of $\rho_{\cal C}$ and $\rho_{th}$ agree in the late time limit.

Now consider a perturbed state $\rho_{\cal C} + \delta\rho$ which also asymptotes to a stationary black hole dual to the thermal state $\tilde\rho_{th}$.  By our assumptions, the difference in the Wald entropy $\delta S_W$ between $\rho_{\cal C}$ and $\rho_{\cal C} + \delta\rho$ at late times is equal to the difference in the Wald entropy between $\rho_{th}$ and $\tilde\rho_{th}$ (calculated at any time, since these black holes are stationary).  We can now apply our theorem and obtain
\begin{align} \label{eq:SWLateTime}
\lim_{T\to\infty} \delta S_W = \delta \left< H_{th} \right> = \delta \ScOP(\rho_{th}),
\end{align}
where $H_{th}$ is the modular Hamiltonian of $\rho_{th}$, $T$ parameterizes a foliation of the collapsed black hole horizon, and $\delta \ScOP(\rho_{th})$ is the difference of the one-point entropy between the two stationary black holes.

Eq.~\eqref{eq:SWLateTime} equates $ \delta \left< H_{th} \right>$ and $\delta\ScOP(\rho_{th})$, but the latter quantity is not the same as $\delta\ScOP(\rho_{\cal C})$.  This is because the one-point functions of $\rho_{\cal C}$ and $\rho_{th}$ only agree at late times.  However the one-point entropy can be generalized to capture only the late time behavior of the black hole.  This generalization was called the future one point entropy $\Ssc$ in~\cite{Kelly:2013aja} and is defined as in~\eqref{eq:OnePointDef} and~\eqref{eq:constraints} with the replacement $D[{\cal A}]\rightarrow D^+[{\cal A}]$.  That is to say, $\Ssc$ is a coarse-grained entropy that constrains the expectation values of local operators in the {\em future} domain of dependence of $\cal A$.  It follows immediately from this definition that $\Ssc$ also satisfies a second law in the sense that $\partial_t \Ssc(\rho_{{\cal A}_t}) \ge 0$, where ${\cal A}_t$ is a foliation of $D[{\cal A}]$.

It follows from our assumption that the one-point functions of $\rho_{\cal C}$ and $\rho_{th}$ only agree at late times (along with an additional assumption that $\Ssc$ is suitably continuous) that
\begin{align}
\lim_{T\to\infty} \delta S_W = \lim_{t\to\infty} \delta \Ssc(\rho_{{\cal C}_t}),
\end{align}
where ${\cal C}_t$ a foliation of the boundary spacetime.  This is the analog of~\eqref{eq:ScOPeqSW} for black holes formed from collapse.  It would be interesting in future work to compare these two quantities at large but finite times $t,T$.

\section{Discussion} \label{sec:Disc}

In this note we have shown that the bulk first law for a class of stationary geometries is dual to the coarse-grained first law associated with the one-point entropy $\ScOP$ and that there exist CFT pure states for which this coarse-graining is necessary for~\eqref{eq:ScOPeqSW} to hold.  Our results imply that $S_W$ is not strictly a measure of entanglement in the CFT.

It remains to ask if our results are unique, i.e. is $\ScOP$ the only coarse-grained entropy which is equal to the Wald entropy to linear order?  The answer turns out to be no, any coarse-grained entropy which fixes the expectation value of the modular Hamiltonian will do the job.  To be definite let $\Sc^{(0)}$ be a coarse-grained entropy that fixes all global charges, in our case the total energy and angular momentum.  Because the first law of entanglement entropy is only sensitive to the change in the expectation value of the modular Hamiltonian, we have $\delta\Sc^{(0)} = \delta \ScOP = \delta S_W$.

However, it is easy to see that $\Sc^{(0)}$ is not equal to $S_W$ beyond linear order.  This is because $\Sc^{(0)}$ is always equal to the Wald entropy of a stationary black hole with given energy and angular momentum, so for generic states the second law requires that $S_W<\Sc^{(0)}$.  This implies that if there exists a coarse-grained entropy which is equal to $S_W$ to all orders it would need to constrain more of the state than just the global charges.  It was argued in~\cite{Kelly:2013aja} that $\ScOP$ is a natural candidate for such a coarse-grained entropy.  See section 4.3 of~\cite{Kelly:2013aja} for a discussion of alternate proposals.






We conclude by discussing the relation of our results to the recent work of~\cite{Balasubramanian:2013lsa,Myers:2014jia,Czech:2014wka,Balasubramanian:2014sra}.  Refs.~\cite{Balasubramanian:2013lsa,Myers:2014jia,Czech:2014wka} developed a formula for computing the area of closed bulk surfaces in terms of a quantity called the differential entropy.  The differential entropy explicitly makes use of locally-extremal (but not necessarily minimal) surfaces.  It was then argued in~\cite{Balasubramanian:2014sra} that non-minimal extremal surfaces in $\text{AdS}_{2+1}$ measure CFT `entwinement', defined as the entanglement entropy between degrees of freedom which are not necessarily spatially localized.  This interpretation refines the proposal of~\cite{Balasubramanian:2013lsa} that the differential entropy measures the information that is not accessible to a family of causal observes in a finite amount of time.

The causal information surface $C_{\cal G}$  (see Fig.~\ref{fig:causal}), where $\cal G$ is a boundary Cauchy surface of the geon spacetime mentioned above, provides in interesting setting for studying these proposals.\footnote{More precisely we are interested in the limit as we approach $C_{\cal G}$ from the black hole exterior.  The quotient used to construct the $\mathbb{RP}_n$ geon introduces an unphysical discontinuity in the area of spheres at $C_{\cal G}$, but the limit is well behaved.}  For this surface, the differential entropy takes a particularly simple form, it is given by the area of a single locally-extremal (but not minimal) surface.  The one-point entropy can also be calculated exactly and agrees with the area of this surface (as predicted by the conjectured formula~\eqref{eq:SeqCHI}).

Curiously, the same surface is singled out by both the differential entropy and $\ScOP$, but for different reasons.  The surface $C_{\cal G}$ is a simple measure of entwinement because it is an extremal surface and it is conjectured to be a measure of the one-point entropy because it lies at the intersection of causal horizons.  It would be interesting to understand how these measures of information are related as the spacetime is perturbed and the extremal and causal surfaces no longer coincide.  Unfortunately, this difference does not show up in our linearized analysis precisely because the surface is extremal and therefore the area is not sensitive to the position of the surface at linear order.  It seems that what is needed are more powerful methods of calculating $\ScOP$ both for testing~\eqref{eq:SeqCHI} and for comparing $\ScOP$ with the differential entropy.

\section*{Acknowledgements}

It is a pleasure to thank Kevin Kuns, Don Marolf and Aron Wall for helpful feedback and discussions.  This work was supported in part by the National Science Foundation under Grant No PHY12-05500, by FQXi grant FRP3-1338, and by funds from the University of California.

\appendix

\section{Testing $\ScOP = \chi$} \label{app:tests}

In this appendix we propose a testable conjecture about the Einstein equation which, if true, could provide substantial evidence for~\eqref{eq:SeqCHI}.  Attempts to carry out these tests are ongoing and will be reported separately.

Let $s$ be a smooth, asymptotically locally AdS solution to the vacuum Einstein equation.  Let $g_{\mu\nu},T^{\mu\nu}$ be the boundary metric and stress tensor of $s$ and let $\cal A$ be some spacelike region on the boundary.  Now let $\cal S$ be the set of all smooth asymptotically AdS solutions $\tilde s$ with boundary data $\tilde g_{\mu\nu},\tilde T^{\mu\nu}$ such there exists a region $\tilde{\cal A}$ on the boundary of $\tilde s$ which satisfies
\begin{align} \label{eq:boundarydata}
g_{\mu\nu}(x) = \tilde g_{\mu\nu}(x),\quad T^{\mu\nu}(x) = {\tilde T}^{\mu\nu}(x), \quad x\in D[\tilde{\cal A}].
\end{align}
These classical solutions $\cal S$ capture some subset of the quantum states $S_{\cal A} \subset T_{\cal A}$ over which we would like to maximize the von Neumann entropy in order to evaluate~\eqref{eq:OnePointDef}.  

Now we introduce a new assumption.  Say that,
\begin{align} \label{eq:assumption}
\underset{\tau_{\cal A} \in T_{\cal A}}{\text{lub}} S(\tau_{\cal A}) = \underset{\sigma_{\cal A} \in S_{\cal A}}{\text{lub}} S(\sigma_{\cal A}).
\end{align}
If this assumption holds we may calculate $\ScOP$ by considering classical geometries only, and maximizing the entropy reduces to maximizing the area of the extremal surface $E_{\cal A}$ (see Fig.~\ref{fig:causal}) over geometries in $\cal S$.  It should be noted that~\eqref{eq:assumption} holds whenever we have to date been able to calculate $\ScOP$ (including the perturbative results established in section~\ref{sec:CG}).  

Assuming~\eqref{eq:assumption}, then the conjecture~\eqref{eq:SeqCHI} makes two predictions about $\cal S$: 
\begin{itemize}
\item \label{item:chipreserving}
every solution $\tilde s\in  {\cal S}$ should satisfy 
$\text{Area}[C_{\cal A}(\tilde s)] = \text{Area}[C_{\cal A}( s)]$ 
, and
\item \label{item:lub}
$\text{Area}[C_{\cal A}(s)] = \underset{\tilde s \in {\cal S} }{\text{lub}} \, \text{Area}[E_{\cal A}(\tilde s)]$.
\end{itemize}
The first claim follows from the fact that~\eqref{eq:SeqCHI} implies that $\chi_{\cal A}$ is a function only of the boundary data in $D[{\cal A}]$ which is being held fixed by~\eqref{eq:boundarydata}.  The second claim is simply a combination of our assumption~\eqref{eq:assumption} and~\eqref{eq:SeqCHI}.
We should note that if the first claim $\text{Area}[C_{\cal A}(\tilde s)] = \text{Area}[C_{\cal A}( s)]$ is true, then it follows from existing results~\cite{Hubeny:2012wa,Wall:2012uf} that $\text{Area}[C_{\cal A}(s)]$ is an upper bound on $\text{Area}[E_{\cal A}(\tilde s)]$ (but not that it is the least upper bound).

These conjectures, even if they are difficult to prove in any generality, can in principle be tested by constructing solutions numerically.  Such tests have the potential to provide strong evidence for (or to conclusively falsify)~\eqref{eq:SeqCHI}.

\bibliographystyle{kp}

\bibliography{FirstLaw8.bbl}

\end{document}